\documentclass{PoS}
\usepackage{aas_macros}

\title{BurstCube: Concept, Performance, and Status}

\ShortTitle{BurstCube}

\author{\speaker{Jacob R. Smith}\\
        NASA/GSFC/CRESST\\
        E-mail: \email{jacob.r.smith@nasa.gov}}
        
\author{
Michael S. Briggs$^a$, 
Alessandro Bruno$^b$, 
Eric Burns$^b$, 
Regina Caputo$^b$, 
Brad Cenko$^b$, 
Antonino Cucchiara$^c$, 
Georgia de Nolfo$^b$, 
Sean Griffin$^d$, 
Lorraine Hanlon$^e$, 
Dieter H. Hartmann$^f$, 
Michelle Hui$^g$, 
Alyson Joens$^h$, 
Carolyn Kierans$^b$, 
Dan Kocevski$^g$, 
John Krizmanic$^d$, 
Amy Lien$^d$, 
Sheila McBreen$^e$, 
Julie E. McEnery$^b$, 
Lee Mitchell$^i$, 
David Morris$^c$, 
David Murphy$^e$, 
Jeremy S. Perkins$^b$, 
Judy Racusin$^b$, 
Peter Shawhan$^j$, 
Teresa Tatoli$^d$, 
Alexey Uliyanov$^e$, 
Sarah Walsh$^e$, 
Colleen Wilson-Hodge$^g$ 
\\
\llap{$^a$}University of Alabama, Huntsville\\
\llap{$^b$}NASA/GSFC\\
\llap{$^c$}University of the Virgin Islands\\
\llap{$^d$}NASA/GSFC/CRESST\\
\llap{$^e$}University College Dublin\\
\llap{$^f$}Clemson University \\
\llap{$^g$}NASA/MSFC\\
\llap{$^h$}George Washington University \\
\llap{$^i$}Naval Research Laboratory\\
\llap{$^j$}University of Maryland, College Park\\
}        
        
\abstract{The first simultaneous detection of a short gamma-ray burst (SGRB) with a gravitational-wave (GW) signal ushered in a new era of multi-messenger astronomy. In order to increase the number of SGRB-GW simultaneous detections, we need full sky coverage in the gamma-ray regime. BurstCube, a CubeSat for Gravitational Wave Counterparts, aims to expand sky coverage in order to detect and localize gamma-ray bursts (GRBs). BurstCube will be comprised of 4 Cesium Iodide scintillators coupled to arrays of Silicon photo-multipliers on a 6U CubeSat bus (a single U corresponds to cubic unit $\sim$10 cm $\times$ 10 cm $\times$ 10 cm) and will be sensitive to gamma-rays between 50 keV and 1 MeV, the ideal energy range for GRB prompt emission. BurstCube will assist current observatories, such as {\it Swift} and {\it Fermi}, in the detection of GRBs as well as provide astronomical context to gravitational wave events detected by Advanced LIGO, Advanced Virgo, and KAGRA. BurstCube is currently in its development and testing phase to prepare for launch readiness in the fall of 2021. We present the mission concept, preliminary performance, and status.}

\FullConference{36th International Cosmic Ray Conference -ICRC2019-\\
		July 24th - August 1st, 2019\\
		Madison, WI, U.S.A.}

\begin{document}

\section{Introduction}
BurstCube detections of short gamma-ray burst (SGRBs) in the multimessenger era will provide critical observations needed to unveil the origins and properties of binary neutron star (BNS) mergers, and possibly neutron star black hole (NSBH) mergers, if discovered. 
Jets that power GRBs are ultra-relativistic, highly-collimated beamed outflows and are some of the shortest-lived and most energetic jets in the universe 
\cite{1986ApJ...308L..47G}.
Electromagnetic follow-up observations of these jet sources revealed their cosmological origins, showed that they have extremely compact central engines and constitute two classes based on emission duration. SGRBs have emission durations less than 2 seconds and long duration burst (LGRBs) last longer than 2 seconds 
\cite{1993ApJ...413L.101K}. 
The merger hypothesis for SGRBs \cite{1992ApJ...395L..83N}
was developed from these observations and spectacularly confirmed with GW170817/GRB17017A 
\cite{PhysRevLett.119.161101, Goldstein_2017, Savchenko_2017}.

GW observations directly probe the central engine, and combining this with information from the electromagetic sector offers additional insights into these events.
Coincident detections enable additional observations from a wealth of multi-wavelength EM telescopes in operation today and the early 2020s \cite{Villar_2017}.
BurstCube observations will provide complementary information to other GRB monitors. From these we will learn about the energetics of the jet, bulk Lorentz factor of the outflow and geometry of the emission region. Time delay between the merger and the SGRB allows us to study jet physics, neutron star equation of state, and the speed of gravity \cite{Abbott_2017}.
Fundamental progress can be achieved for NS mergers and GRB jet physics with the combined observations of the GW network, current GRB monitors, and BurstCube.

\section{BurstCube}
Thanks to the growth and development of small-sat and CubeSat technologies it is now achievable to build a CubeSat to detect the bright signals from GRBs. These proceedings detail a 6U CubeSat (a single U corresponds to cubic unit $\sim$10 cm $\times$ 10 cm $\times$ 10 cm)  \cite{cubesat}
mission called BurstCube that is based on the Dellingr platform developed at GSFC
\cite{2015EPSC...10..720J}. 
The main limiting factors in developing a mission like BurstCube are mass, power, and volume but the advent of low-power, low-volume SiPMs enables a workable detector readout design. We describe the BurstCube CubeSat, which is designed to provide rapidly available high-resolution temporal, spectral, and localization data. Even rough localizations will be helpful to verify weak GW signals, as the temporal coincidence is the most important measurement alerting the community to a potential on-axis event. BurstCube increases the sky coverage of existing facilities, and is optimized for detecting SGRBs, all for a small
fraction of the cost of a larger mission.

\begin{figure}
\centering
\begin{minipage}{0.7\textwidth}
\includegraphics[width=\textwidth,trim=0 0 0 0,clip=true]{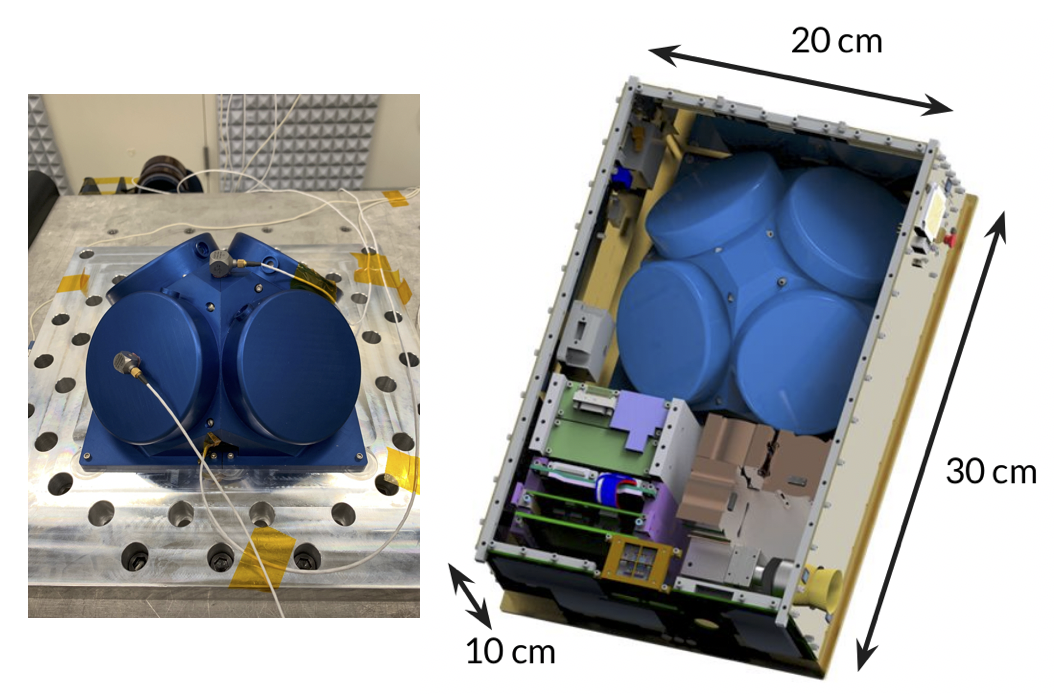}
\end{minipage}\hfill
\begin{minipage}{0.3\textwidth}
\caption{\small Internal view of the BurstCube instrument and spacecraft components ({\it right}), and the instrument mounted on a vibration (vibe) table ({\it left}). For the vibe test, one quarter was instrumented with a CsI(Tl) crystal and a proto-flight SiPM array front-end electronics board and the other three with mock masses. Tests consisted of flight-like vibe amplitudes on all three axes. BurstCube vibe was completed and successfully passed in June 2019. 
\label{fig:spacecraft}}
\end{minipage}
\end{figure}

\begin{figure}
\centering
\includegraphics[width=0.8\textwidth,trim=0 0 0 0,clip=true]{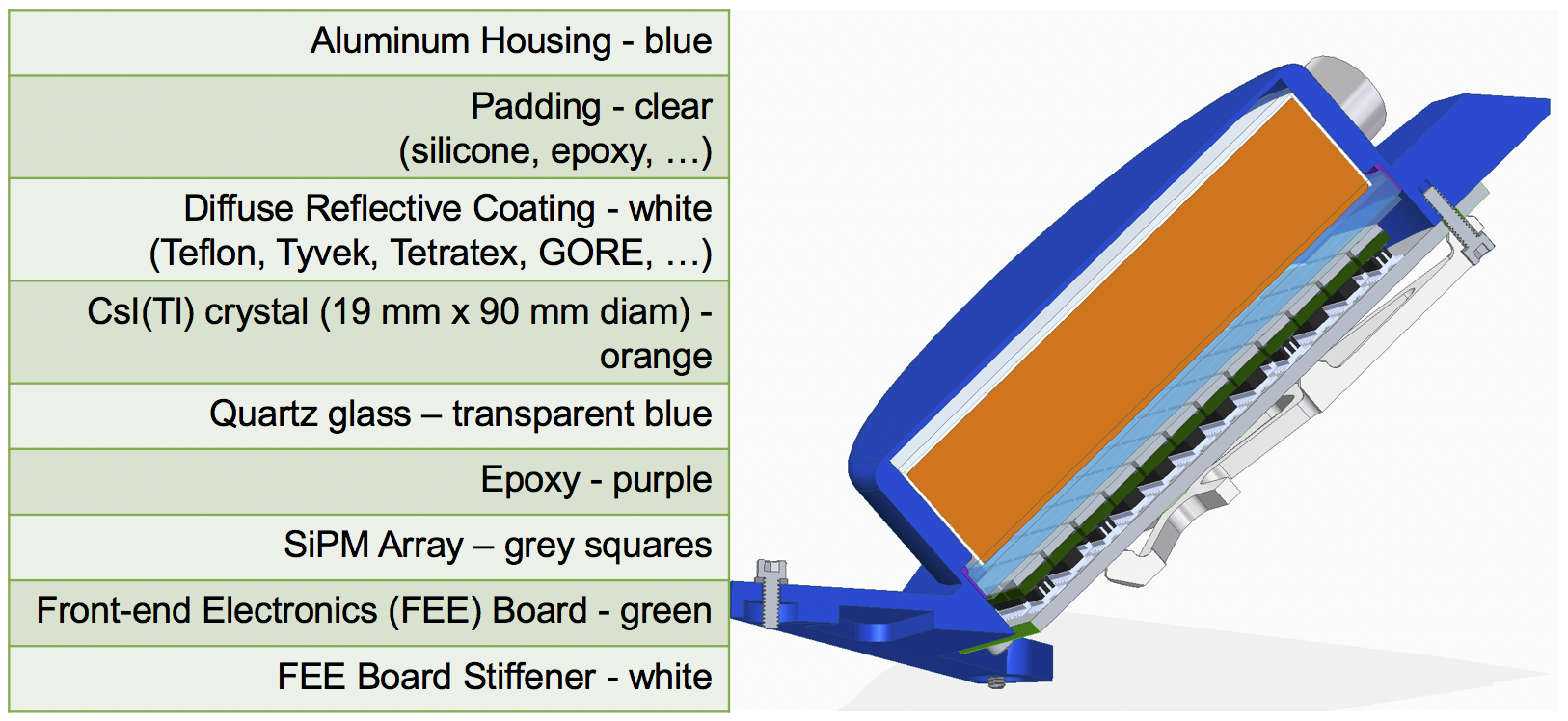}
\caption{\small Cross-section view of the BurstCube SQD ({\it right}) with a description of the internal components ({\it left}).
\label{fig:sqd}}
\end{figure}

\begin{figure}
\centering
\begin{minipage}{0.52\textwidth}
\centering
\includegraphics[width=0.8\textwidth,trim=0 0 0 0,clip=true]{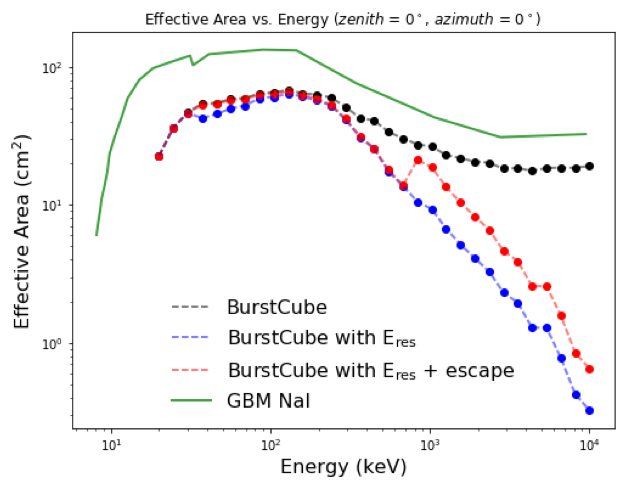}
\caption{\small Despite the constraints of a CubeSat, detector simulations show that BurstCube achieves an effective area
of 70\% of GBM at 100 keV and $15^\circ$ incidence 
The effective area as a function of energy, and the corresponding
curve for the larger GBM NaI detectors are shown for reference.\label{fig:effarea_energy}}
\end{minipage}\hfill
\begin{minipage}{0.44\textwidth}
\includegraphics[width=1.05\textwidth,trim=0 0 0 0,clip=true]{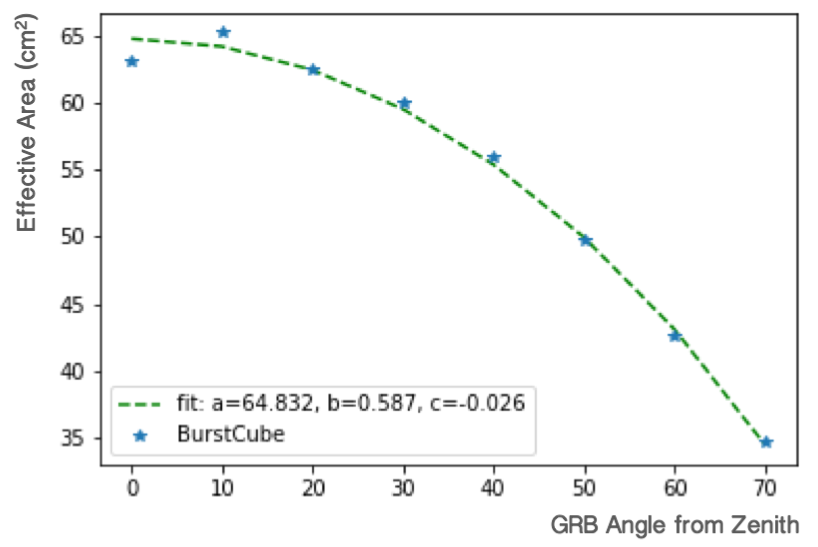}
\caption{\small The cylindrical BurstCube detectors make maximum use of
  space and still achieve an effective area cosine dependence on
  incidence angle expected for a thin detector which is needed for GRB localizations.
\label{fig:effarea_angle}}
\end{minipage}
\end{figure}

BurstCube is a 6U CubeSat divided into a 4U instrument package, and 2U of spacecraft subsystems ({\bf Fig. \ref{fig:spacecraft}}). The format of the 6U spacecraft drives the instrument mass, power, and volume budget. The instrument design is similar to
{\it Fermi}-GBM \cite{2009ApJ...702..791M}, except BurstCube uses Thallium doped Cesium Iodide (CsI(Tl)) crystal as the detection medium covering the energy range from 50 keV - 1 MeV with high efficiency and adequate energy resolution. These crystals are inexpensive and have a long track record of use in $\gamma$-ray astronomy. A design philosophy is to rely on proven well-tested technology. The dimensions of the CsI employed for BurstCube are 9 cm in diameter and 1.9 cm thick, with the maximum dimension dictated by the size of the 6U CubeSat.  Readout of each crystal is done by an array of 116 low-power and low-voltage (56-60 V) Hammamatsu Silicon Photomultipliers (SiPMs). Compared to the more conventional photomultiplier tubes, SiPMs significantly reduce mass, volume, power, and cost. However, even in this compact design, the detector presented here is competitive with the state of the art system (\textit{Fermi}-GBM). A single BurstCube crystal has ~70$\%$ of the effective area of a single GBM crystal at 100 keV; see {\bf Fig. \ref{fig:effarea_energy}}. However, GBM has better sky-uniformity and better localizations. The 4U instrument package consists of four Single Quarter Detectors (SQD) each with a PTFE wrapped crystal housed inside an Al structure with quartz glass and silicone optical padding between the crystal and a SiPM front-end electronics (FEE) board, see {\bf Fig. \ref{fig:sqd}}. The combination of scintillation crystals and new readout devices makes it possible to consider a compact, low-power instrument that is readily deployable on a CubeSat platform.
 
To estimate the BurstCube effective area and resulting sensitivity, we simulated a single CsI detector using MEGAlib \cite{Megalib} (which provides an interface to GEANT4). The simulated BurstCube effective area is complementary to GBM ({\bf
  Fig. \ref{fig:effarea_energy}}), despite the smaller detectors. BurstCube will localize GRBs in a similar manner
to BATSE or GBM 
\cite{batse, GBM_GRB_Locs}. 
The approximate cosine dependence ($\sim$cos$(\theta)^{0.59}$) of the
effective area with incidence angle ({\bf Fig.
  \ref{fig:effarea_angle}}) is leveraged to localize GRBs by measuring
the relative brightness of the burst between the four detectors. The detector count rates will be matched to a previously computed table of
relative detector responses as a function of spacecraft coordinates.
For 90\% of SGRBs we are expecting localizations better than $30^\circ$ radius.

The BurstCube instrument is housed in a CubeSat bus based on the Dellingr platform,  a modular platform designed to be easily modifiable for a variety of 6U CubeSat architectures (see {\bf Fig. \ref{fig:spacecraft}}). It includes 3 axis-stabilized pointing control, star tracker, deployable mounted solar panels, and is designed for the Planetary Systems Corporation (PSC) 6U canister standard\footnote{Also compatible with the NanoRacks 6U dispenser.} \cite{pscCubesat}.  A CubeSat implementation of the Vulcan communications system is currently being developed for use in BurstCube. The Vulcan radio will enable rapid GRB location distribution through the Tracking and Data Relay Satellite System (TDRSS).  

\section{Performance}

\begin{figure}
\centering
\centering
\includegraphics[width=1.0\textwidth,trim=0 0 0 0,clip=true]{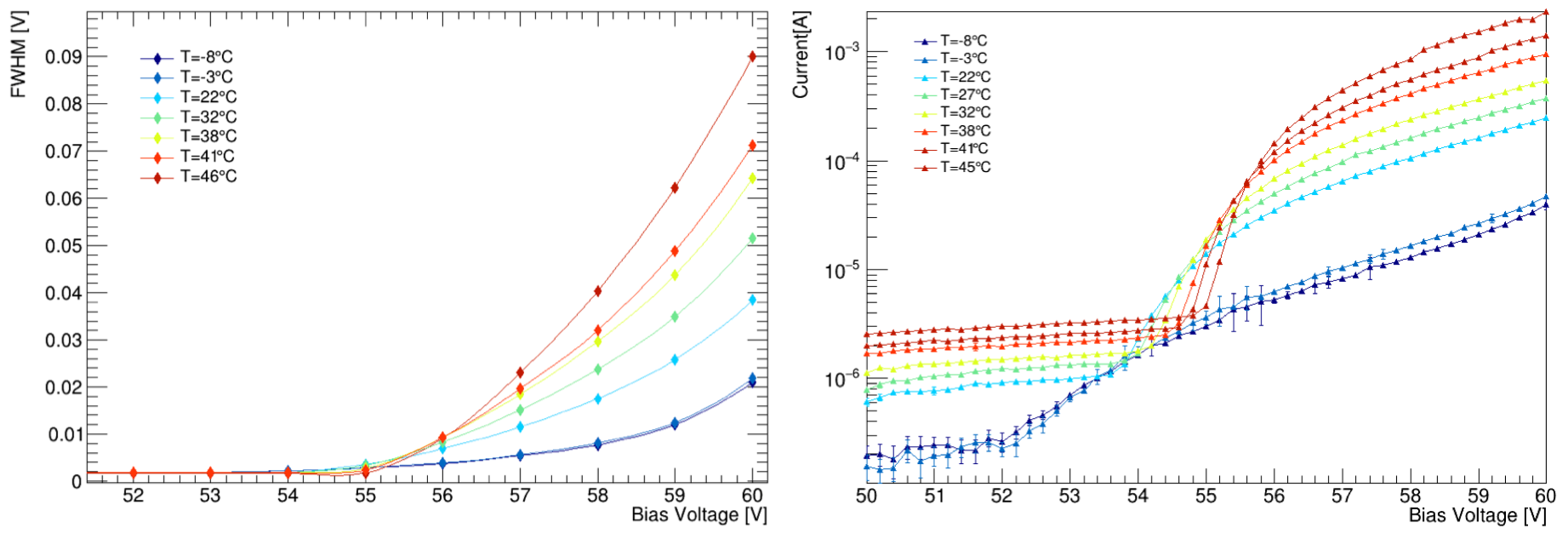}
\caption{
\small 
Noise of the SiPM array ({\it left}), represented as FWHM [V] as a function of bias voltage, and breakdown voltage ({\it right}), represented in terms of an I-V curve as a function of bias voltage, was measured with the BurstCube proto-flight SiPM FEE board. 
BurstCube can detect down to 26 keV and achieve constant gain through temperature compensation within its expected flight temperature environment.
\label{fig:sipm_noise_iv}}

\end{figure}

In addition to detector simulations, expected performance of the BurstCube mission has been determined from laboratory measurements using proto-flight components of the instrument. In the last year we have constructed a proto-flight of the instrument that consists of a front-end electronics (FEE) board and a SQD. The FEE board is where the array of 116 SiPMs and associated summing electronics is installed. We have characterized the temperature response, noise behavior, and breakdown voltage of the SiPM array with the use of various gamma-ray radiation sources in the energy range from 26 keV up to 1.33 MeV. Most of the characterization of the BurstCube SiPM FEE board has been done using a 10 cm diameter, 2 cm thick CsI(Na) test crystal.  {\bf Fig. \ref{fig:sipm_noise_iv}} shows our measured noise response and I-V curve, representing the breakdown voltage, of the SiPM FEE as a function of bias voltage and temperature. Not shown here is the energy response to the 26 keV and 59 keV gamma-ray emission from an Americium-241 radiation source. The 26 keV photo-peak is distinguishable from noise below 38 C and the 59 keV photo-peak is distinguishable from noise below 41 C. Based on initial orbital temperature analysis, the hottest expected temperature of BurstCube is below 35 C. These results indicate that BurstCube will detect gamma-rays below its low energy goal of 50 keV. 

The breakdown voltage of SiPMs have a strong dependence on temperature. During the flight operations, BurstCube will experience a range of temperatures between 5--35 $^\circ$C. As a result we need to demonstrate an ability to compensate for these fluctuations in temperature. Shown in right panel of {\bf Fig. \ref{fig:sipm_noise_iv}} are lab measurements of the I-V curve over a range of temperatures. The breakdown voltage at each temperature is indicated by the sudden increase of current at a particular bias voltage. An on-board look-up-table (LUT) will be used in flight to compensate the temperature variations by adjusting the applied bias voltage. In this way, BurstCube can maintain constant gain within the temperature fluctations of the mission lifetime. 

\begin{figure}
\centering
\includegraphics[width=0.75\textwidth,trim=0 0 0 0,clip=true]{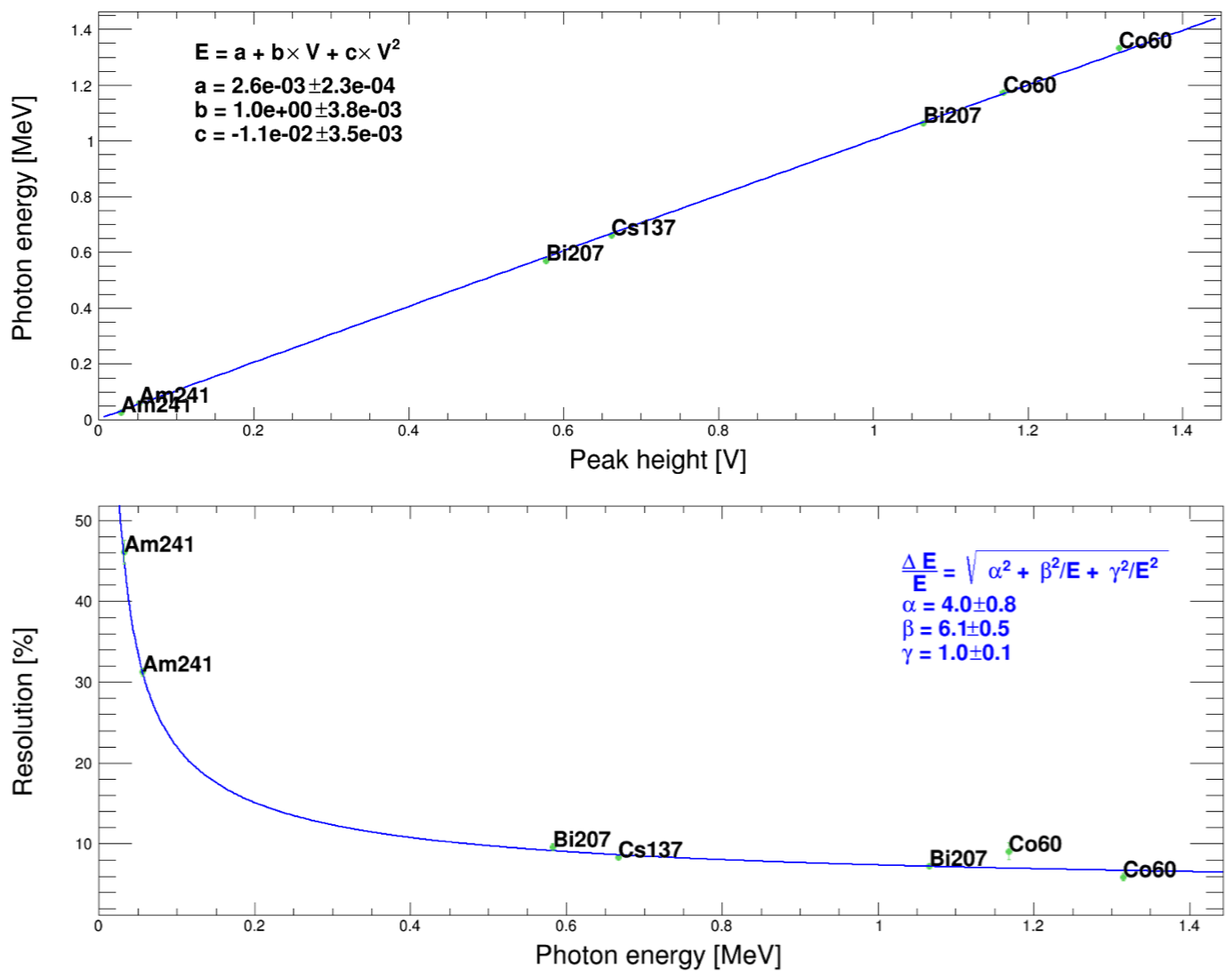}
\caption{\small Energy linearity ({\it top}) and resolution ({\it bottom}) for the BurstCube SiPM FEE with the test crystal using gamma-ray radiation sources between 26 keV up to 1.33 MeV. The SiPM array is linear and has sufficient energy resolution over the desired energy range of the mission.
\label{fig:energyResponse}}
\end{figure}

Data have been collected with the BurstCube SiPM FEE board with radiation sources including Am-241, Bi-207, Cs-137 and Co-60 to measure the energy response from 26 keV up to 1.33 MeV with a test crystal at room temperature, see {\bf Fig. \ref{fig:energyResponse}}. BurstCube shows a linear response and sufficient energy resolution over the desired energy range of the mission. Given the measured breakdown voltage over the expected temperature range we expect to have constant linear gain and good resolution with the BurstCube 116 SiPM array.

\section{Current Status}
BurstCube is currently approaching the Critical Design Review (CDR) for the instrument detectors and readout and the Systems Peer Review (SPR) for the spacecraft bus components. The Critical Design Review for BurstCube will verify the final design to initiate the flight instrument build. The spacecraft SPR will serve as the decision point to finalize the components and interfaces within the spacecraft bus and components. Key milestones that have been completed leading up to the instrument CDR include: 
{\bf
design, integration and test of the 116 array SiPM front-end electronics (FEE) board;
design of the instrument power supply;
construction of a proto-flight SQD;
completion of the proto-flight environmental qualification vibration and thermal vacuum tests.
}

Environmental qualification of the SQD validates the BurstCube flight instrument design. Following General Environmental Verification Standard (GEVS) for Goddard Space Flight Center (GSFC) flight programs and projects \cite{GEVS}, in July 2019, we have demonstrated by test that BurstCube will satisactorily perform in its mission environment. GEVS provides the requirements test levels for vibration and thermal vacuum tests of flight or flight-like hardware. The energy response of BurstCube to gamma-ray radiation source was used to demonstrate BurstCube's performance before, during and after environmental tests. Constant gain, linear energy response, and energy resolution was maintained after vibration tests and during thermal vacuum cycles. During the temperature cycles of thermal vacuum tests we were able to maintain constant electronics gain and energy response by compensating the bias voltage of the 116 array SiPM FEE board. Following an independent assessment on the technology readiness level (TRL), the BurstCube instrument will be at TRL 6.

Between the successful completion of the environmental tests and CDR, we plan to: build a flat-sat of the instrument digital electronics with an FPGA and flight-like bias power supply; test FPGA code with instrument and spacecraft interfaces; test FPGA trigger, data processing, and temperature compensation codes; finalize the requirements of the instrument flight software; finalize requirements for the ground pipelines and science analysis software; and complete the full instrument detector simulations. Calibration test plans are also currently in the draft stage that will be used for the flight build of BurstCube.

\bibliographystyle{JHEP}

\providecommand{\href}[2]{#2}\begingroup\raggedright\begin{thebibliography}{10}

\bibitem{1986ApJ...308L..47G}
J.~{Goodman}, \emph{{Are gamma-ray bursts optically thick?}},
  \href{http://dx.doi.org/10.1086/184741}{\emph{\apjl} {\bfseries 308} (Sept.,
  1986) L47--L50}.

\bibitem{1993ApJ...413L.101K}
C.~{Kouveliotou}, C.~A. {Meegan}, G.~J. {Fishman}, N.~P. {Bhat}, M.~S.
  {Briggs}, T.~M. {Koshut} et~al., \emph{{Identification of two classes of
  gamma-ray bursts}}, \href{http://dx.doi.org/10.1086/186969}{\emph{\apjl}
  {\bfseries 413} (Aug., 1993) L101--L104}.

\bibitem{1992ApJ...395L..83N}
R.~{Narayan}, B.~{Paczynski} and T.~{Piran}, \emph{{Gamma-ray bursts as the
  death throes of massive binary stars}},
  \href{http://dx.doi.org/10.1086/186493}{\emph{\apjl} {\bfseries 395} (Aug.,
  1992) L83--L86}, [\href{https://arxiv.org/abs/astro-ph/9204001}{{\ttfamily
  astro-ph/9204001}}].

\bibitem{PhysRevLett.119.161101}
{\scshape LIGO Scientific Collaboration and Virgo Collaboration} collaboration,
  B.~P. Abbott, R.~Abbott, T.~D. Abbott, F.~Acernese, K.~Ackley, C.~Adams
  et~al., \emph{Gw170817: Observation of gravitational waves from a binary
  neutron star inspiral},
  \href{http://dx.doi.org/10.1103/PhysRevLett.119.161101}{\emph{Phys. Rev.
  Lett.} {\bfseries 119} (Oct, 2017) 161101}.

\bibitem{Goldstein_2017}
A.~Goldstein, P.~Veres, E.~Burns, M.~S. Briggs, R.~Hamburg, D.~Kocevski et~al.,
  \emph{An ordinary short gamma-ray burst with extraordinary implications:
  Fermi-{GBM} detection of {GRB} 170817a},
  \href{http://dx.doi.org/10.3847/2041-8213/aa8f41}{\emph{The Astrophysical
  Journal} {\bfseries 848} (oct, 2017) L14}.

\bibitem{Savchenko_2017}
V.~Savchenko, C.~Ferrigno, E.~Kuulkers, A.~Bazzano, E.~Bozzo, S.~Brandt et~al.,
  \emph{{INTEGRAL} detection of the first prompt gamma-ray signal coincident
  with the gravitational-wave event {GW}170817},
  \href{http://dx.doi.org/10.3847/2041-8213/aa8f94}{\emph{The Astrophysical
  Journal} {\bfseries 848} (oct, 2017) L15}.

\bibitem{Villar_2017}
V.~A. Villar, J.~Guillochon, E.~Berger, B.~D. Metzger, P.~S. Cowperthwaite,
  M.~Nicholl et~al., \emph{The combined ultraviolet, optical, and near-infrared
  light curves of the kilonova associated with the binary neutron star merger
  {GW}170817: Unified data set, analytic models, and physical implications},
  \href{http://dx.doi.org/10.3847/2041-8213/aa9c84}{\emph{The Astrophysical
  Journal} {\bfseries 851} (dec, 2017) L21}.

\bibitem{Abbott_2017}
B.~P. Abbott, R.~Abbott, T.~D. Abbott, F.~Acernese, K.~Ackley, C.~Adams et~al.,
  \emph{Gravitational waves and gamma-rays from a binary neutron star merger:
  {GW}170817 and {GRB} 170817a},
  \href{http://dx.doi.org/10.3847/2041-8213/aa920c}{\emph{The Astrophysical
  Journal} {\bfseries 848} (oct, 2017) L13}.

\bibitem{cubesat}
``Cubesat design specification rev. 13.'' The CubeSat Program, Cal Poly SLO.

\bibitem{2015EPSC...10..720J}
M.~{Johnson}, T.~{Bonalsky}, D.~{Chornay}, C.~{Clagett}, A.~{Cudmore},
  A.~{Ericsson} et~al., \emph{{Dellingr- A Path to Compelling Science with
  CubeSats}},  in \emph{European Planetary Science Congress},
  pp.~EPSC2015--720, Oct, 2015.

\bibitem{2009ApJ...702..791M}
C.~{Meegan}
  et~al.\href{http://dx.doi.org/10.1088/0004-637X/702/1/791}{\emph{\apj}
  {\bfseries 702} (Sept., 2009) 791--804},
  [\href{https://arxiv.org/abs/0908.0450}{{\ttfamily 0908.0450}}].

\bibitem{Megalib}
A.~{Zoglauer}, R.~{Andritschke} and F.~{Schopper}, \emph{"megalib the medium
  energy gamma-ray astronomy library"}, .

\bibitem{batse}
G.~J. {Fishman} et~al., \emph{{Overview of Observations from BATSE on the
  Compton Observatory}}, {\emph{\aaps} {\bfseries 97} (Jan., 1993) 17}.

\bibitem{GBM_GRB_Locs}
V.~{Connaughton} et~al., \emph{{Localization of Gamma-Ray Bursts Using the
  Fermi Gamma-Ray Burst Monitor}},
  \href{http://dx.doi.org/10.1088/0067-0049/216/2/32}{\emph{\apjs} {\bfseries
  216} (Feb., 2015) 32}, [\href{https://arxiv.org/abs/1411.2685}{{\ttfamily
  1411.2685}}].

\bibitem{pscCubesat}
``Canisterized satellite dispenser data sheet.'' 2002337F.

\bibitem{GEVS}
``General environmental verification standard (gevs) for gsfc flight programs
  and projects.'' GSFC-STD-7000A, 2013.

\end{thebibliography}\endgroup

\providecommand{\href}[2]{#2}\begingroup\raggedright\endgroup

\end{document}